\documentclass[english]{article}
\usepackage[T1]{fontenc}
\usepackage[latin9]{inputenc}
\usepackage{float}
\usepackage{textcomp}
\usepackage{amsthm}
\usepackage{amsmath}
\usepackage{graphicx}
\usepackage{amssymb}
\usepackage[numbers]{natbib}

\makeatletter

\providecommand{\tabularnewline}{\\}

\makeatother

\usepackage{babel}

\begin{document}

\title{About functions where function input describes inner working of the
function}

\author{Rade Vuckovac rade.vuckovac@gmail.com}
\maketitle
\begin{abstract}
This paper argues an existence of a class of functions where function
own input makes function description. That fact have impact to the
wide spectrum of phenomena such as negative findings of Random Oracle
Model in cryptography, complexity in some rules of cellular automata
(Wolfram rule 30) and determinism in the true randomness to name just
a few.
\end{abstract}

\section{Introduction}

This paper is a continuation of research done on MAG type algorithms
\citep{key-1}. Although MAG algorithms are based on the use of the
McCabe conditional complexity, the underlying complexity mechanism
was never explored further. In this attempt we try to merge algorithms
and composite functions, and also we point out the missing ingredient
in defining a function\textquoteright{}s description.

\section{The Mechanism behind complexity in systems with high conditional
complexity}

\subsection{Conditional complexity}

The McCabe \citep{key-3} work on software testing is very relevant
here, because it relates graph theory and algorithm input / output.
The essence of this work is to treat an algorithm branching as a tree
structure introducing the phenomena of conditional complexity. The
conditional complexity is a metric or a number of independent paths
through some software module. For example, if a source code of an
algorithm does not have if / else algorithm structure then there is
only one path through the source code. If there is one if / else then
there are two paths through source code, one path when the conditional
argument is TRUE and the other when the conditional argument is FALSE.
See the figure 1.

\begin{figure}
\caption{\protect\includegraphics[scale=3]{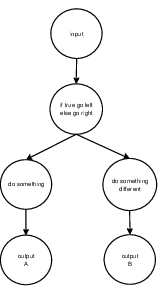}}

\end{figure}

Consequently, the if / else statement will double the number of possible
paths through the source code, making a software module more complex.
Therefore there is potential for exponentially more independent output
cases for testing. 

It is an interesting observation of McCabe that small programs (in
lines of code terms) usually make complex ones because of the inclusion
of several or more if/else programming compositions.

\subsection{An algorithm as a function}

The whole idea of conditional complexity can be taken a step further.
Usually a function is given by a formula or a plot, or it is computed
by an algorithm. The figure 2 from Wikipedia gives a general idea
of a function. The next logical step is to treat an algorithm as a
function, not just as a tool to compute a function. Certainly an algorithm
can be a function in own right. For example, let us consider the algorithm
from figure 3. Like a function, here we have an input and an output.
The \textquotedblleft{}Black box\textquotedblright{} of the function
in principle can be anything; in this case it is an algorithm (the
same idea as figure 2).

\begin{figure}

\caption{\protect\includegraphics[scale=3]{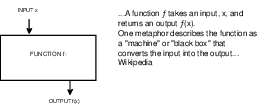}}

\end{figure}

\begin{figure}[h]
\caption{\protect\includegraphics[scale=3]{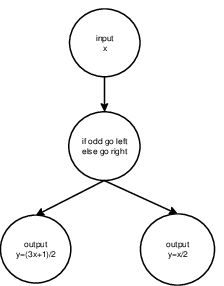}}

\end{figure}

\subsection{3n+1 problem (the Collatz conjecture)\citep{key-2}}
\begin{quotation}
... The Collatz conjecture is an unsolved conjecture in mathematics.
It is named after Lothar Collatz, who first proposed it in 1937. The
conjecture is also known as the 3n + 1 conjecture, as the Ulam conjecture
(after Stanislaw Ulam), or as the Syracuse problem; the sequence of
numbers involved is referred to as the hailstone sequence or hailstone
numbers, or as wondrous numbers per Gödel, Escher, Bach. We take any
number n (element of N+). If n is even, we halve it (n/2), else we
do \textquotedbl{}triple plus one\textquotedbl{} and get 3n+1. The
conjecture is that for all numbers this process converges to 1. Hence
it has been called \textquotedbl{}Half Or Triple Plus One\textquotedbl{},
sometimes called HOTPO. Paul Erd\H{o}s said about the Collatz conjecture:
\textquotedbl{}Mathematics is not yet ready for such problems.\textquotedbl{}
He offered \$500 for its solution. (Lagarias 1985) ... 

Experimental evidence The conjecture has been checked by computer
for all starting values up to 10 \texttimes{} 2\textasciicircum{}58 \ensuremath{\approx} 2.88\texttimes{}10\textasciicircum{}18.
While impressive, such computer evidence should be interpreted cautiously.
More than one important conjecture has been found false, but only
with very large counterexamples. (See for example the Pólya conjecture,
the Mertens conjecture and the Skewes' number.) ... (Wikipedia) 
\end{quotation}
\begin{figure}

\caption{\protect\includegraphics[scale=3]{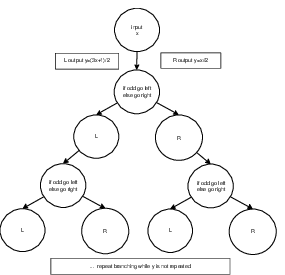}}

\end{figure}

Interestingly enough our figure 3 with little massaging is exactly
the 3n+1 problem. (see figure 4). It should be noted that this algorithm
does not stop necessarily on y = 1. It will stop when the y value
is repeated or in other case it will branch forever (which is also
interesting prospect).\textbf{\textit{ Here we can say that there
is the set of numbers of all checked numbers so far (up to $10\times2^{58}$ )
that obeys the Collatz conjecture and we will call that set S}}. Also
we observe that all members of that set are mapped to 1 by the algorithm
/ function as shown in figure 4. Now our figure 4 almost resembles
the diagram of a composite function as shown in figure 5. 

\begin{figure}
\caption{\protect\includegraphics[scale=3]{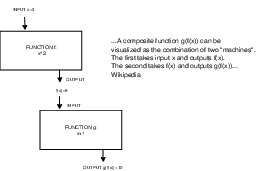}}

\end{figure}

\textbf{\textit{The fundamental difference between figure 4 and 5
is that a composite function defines strictly one path only. That
is a contrast to our figure 4 which leaves the composition of a particular
path open.}} In both cases, the processes are deterministic. But in
the case of figure 4 the algorithm / function is not fully described.
According to the conditional complexity the input may went through
any possible path towards the output. And if we enumerate left branching
as f and right branching as g, for first two levels of branching in
figure 4 we have 4 combinations of f and g:

\begin{tabular}{|c|}
\hline 
combinations f,g\tabularnewline
\hline
\hline 
$fof$\tabularnewline
\hline 
$fog$\tabularnewline
\hline 
$gog$\tabularnewline
\hline 
$gof$\tabularnewline
\hline
\end{tabular}

We also know that order of execution is important because if we start
with function g instead of f (figure 5) we will have different output
(meaning different mapping). That means we have 4 distinct mapping
for first two levels of branching (figure 4) and depending on input
value one of 4 will be chosen for {}``input to output'' transformation.
It can be noted that number of unique arrangements of f\textquoteright{}s
and g\textquoteright{}s will exponentially rise with every consequent
branching. The empirical evidence of Collatz conjecture confirms the
uniqueness of every mapping because for every known input the outcome
will converge to 1.

To fully describe the algorithm / function from figure 4, i.e. to
enable determination of the argument though calculation, every conditional
decision must be recorded: for instance every left turn with \textquoteleft{}L\textquoteright{}
and every right turn with \textquoteleft{}R\textquoteright{}. The
encoding of the description of the particular mapping thus will be
a string of L\textquoteright{}s and R\textquoteright{}s. That string
can serve as a full description of particular composite function for
particular input (to map amount and order of f\textquoteright{}s and
g\textquoteright{}s. For two levels of branching from figure 4 the
four possible strings would be:

\begin{tabular}{|c|c|}
\hline 
combinations f,g & makes\tabularnewline
\hline
\hline 
$fof$ & LL\tabularnewline
\hline 
$fog$ & LR\tabularnewline
\hline 
$gog$ & RR\tabularnewline
\hline 
$gof$ & RL\tabularnewline
\hline
\end{tabular}

The same description (L\textquoteright{}s R\textquoteright{}s string)
can be used to recompose the inverse steps through the algorithm (stepping
from y to x). 

Now we come to the important question of a function\textquoteright{}s
description. Algorithmic information theory, and particularly the
Kolmogorov \textendash{} Chaitin complexity paradigm\citep{key-4}
will use the algorithm from figure 4 as a measurement of complexity
and therefore decide that the complexity of mapping is relatively
low, because whole algorithm can be coded in tens lines of code. On
the other hand there is considerable difficulty (\textquotedbl{}Mathematics
is not yet ready for such problems.\textquotedbl{} Paul Erd\H{o}s)
to find any formal way to predict the function output of figure 4. 

To illustrate a \textquotedblleft{}prediction quality\textquotedblright{}
of classical functions figure 6 can be used. We can see that b is
dependent on angle, so we can say with certainty that if 'a' is rotated
in clock wise direction (making angle less than 60 degrees but still
more than 0 degrees) then 'b' will lay somewhere between 50 and 100.
In contrast, for the case of the figure 4 we have $10\times2^{58}$ inputs
converging to 1, but there is no way excluding intuition (in other
words there is no formal way) of predicting the output of non-tested
input. That simply means that if we really want to fully describe
the 3n+1 function we have to include the description of every input
path transformation (L\textquoteright{}s R\textquoteright{}s strings). 

That also means, for some cases the Kolmogorov \textendash{} Chaitin
complexity is not entirely accurate because the algorithm description
does not mean full description of a phenomena as in figure 6 but may
be more as in figure 4.

\begin{figure}

\caption{\protect\includegraphics{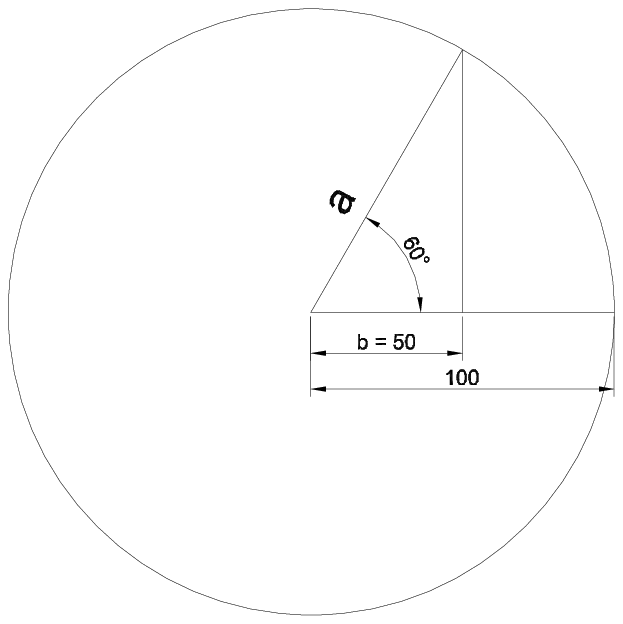}}

\end{figure}

From a skeptical point of view it can always be argued that set of
recorded L\textquoteright{}s R\textquoteright{}s strings may be compressed
in some way (perhaps finding some pattern) and still keep the relatively
low complexity of 3n+1 function\textquoteright{}s description. If
we assume that this is the case, then our L\textquoteright{}s R\textquoteright{}s
descriptions for every input may be shorter than required binary representation
for all inputs. That is unlikely, because L\textquoteright{}s R\textquoteright{}s
descriptions entropy can not be smaller than Shannon's entropy. 

Let say that b (in our case b = 58) is the number of bits required
to describe every input of the set S and r is the number of bits required
to describe all L\textquoteright{}s R\textquoteright{}s recorded strings
for the set S. Because our L\textquoteright{}s R\textquoteright{}s
string is in the same time the description of every path and consequently
the encoding of every input, we can deduce that:

$b\leqq r$

\textbf{That means that the complexity of the full function description
(figure 4, set S) can not be reduced below the complexity of input
(set S).}

It should be noted that there were attempts to prove that the Collatz
conjecture is unprovable (see \citep{key-5}). While argumentation
about unprovable concerns the set of all natural numbers, in this
paper argumentation are within our set S (empirically checked numbers
converging to 1).

\subsection{Stephen Wolfram\textquoteright{}s rule 30}

The rule 30 belongs to class 3 behavior cellular automata (CA). This
class has a complex structure and can be regarded as a chaotic/random
class.

Rule 30 {[}p869 \citep{key-6}{]} can be expressed graphically as
figure 7. The English formulation is {[}p27 \citep{key-6}{]}: \textquotedblleft{}First,
look at each cell and its right-hand neighbor. If both of these were
white on the previous step, then take the new colour of the cell to
be whatever the previous colour of its left-hand neighbor was. Otherwise,
take the new colour to be the opposite of that.\textquotedblright{}

It is obvious that all complexity obtained by this algorithm is a
consequence of the McCabe conditional complexity which is evident
from the above rule (if something is TRUE do something, else do something
different, and the output is next input for the repetition of if /
else, very similar to the figure 4). 

There is also an explanation for the same apparent randomness outcome
(for rule 30) for a start with high entropy and for a start with low
entropy (see figure 8{[}figure 8 \citep{key-6} p281{]}).

The author rightly argues that initial conditions does not play a
role in randomness development for rule 30, opposite to classical
chaotic systems in which small changes in the initial conditions may
strongly effect the outcome. However, the author does not explain
cause of apparent randomness, but now we know that conditional complexity
is indifferent to input entropy and depends solely on the amount of
branching: after so many branching steps the second picture in figure
8 ought to look as the first one after a while.

\begin{figure}[H]
\caption{\protect\includegraphics[scale=0.4]{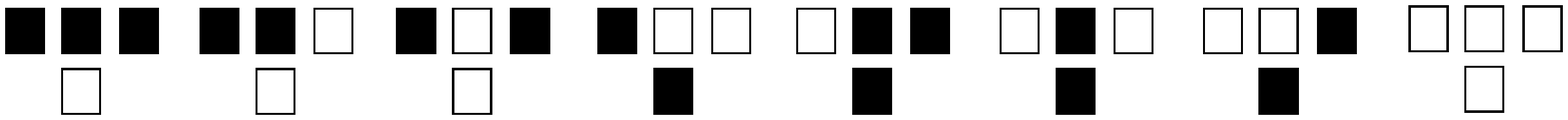}}

\end{figure}

\begin{figure}[H]
\caption{\protect\includegraphics[scale=0.2]{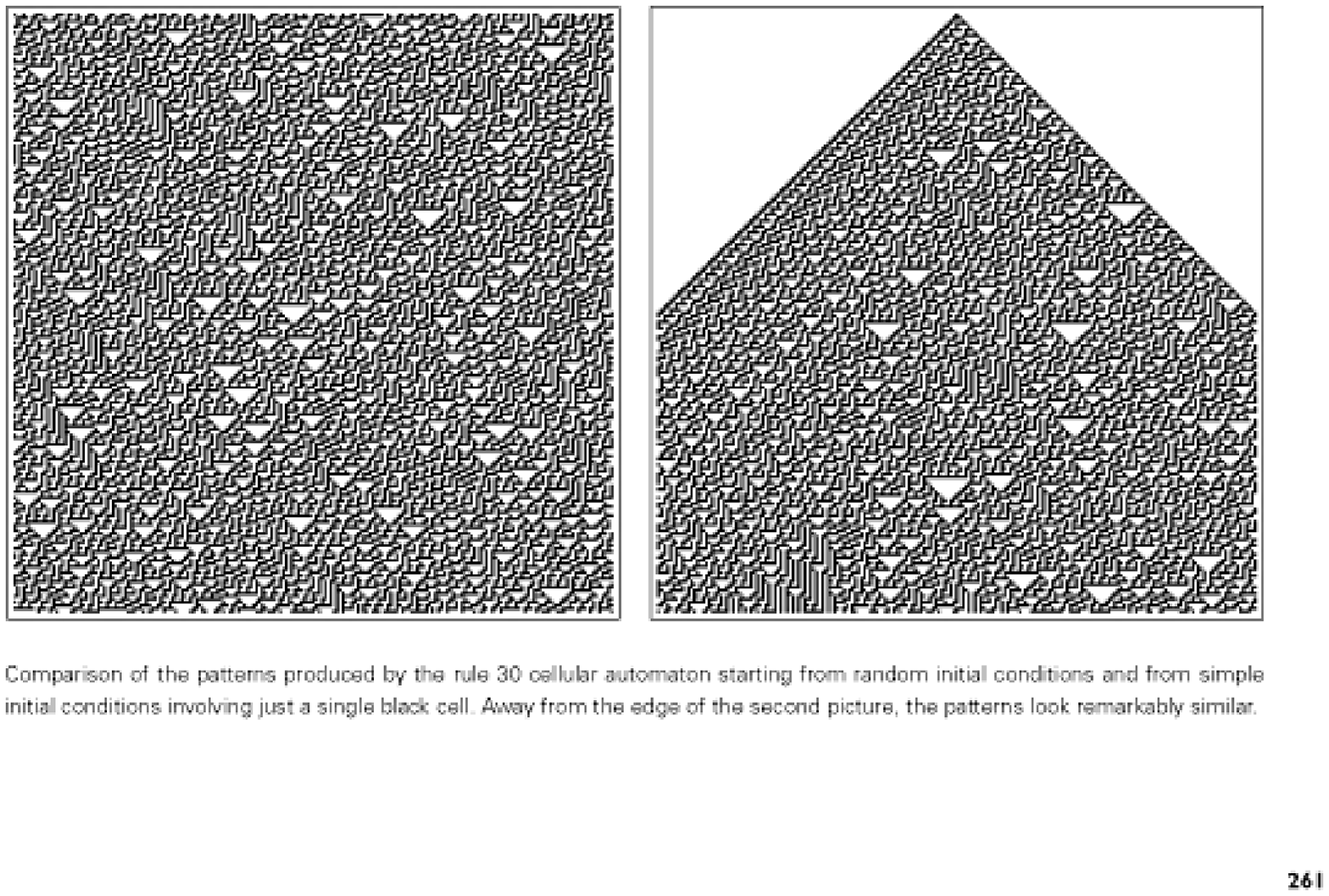}}

\end{figure}

\subsection{Random Oracle Methodology}

Our current understanding of randomness also may be revised. For example
we have statements like \textquotedbl{}Anyone who considers arithmetical
methods of random digits is, of course, in a state of sin.\textquotedbl{}\citep{key-7}
. It will look benign if exchange 'arithmetical methods' with 'deterministic
methods' because of our perception on randomness, but then above statement
is not correct.

Loosely, the formal notation of the above statement is developed through
Random Oracle Methodology(ROM)\citep{key-8} and the methodology\textquoteright{}s
negative results claim that the concrete hash function cannot be substituted
for the random oracle. A problem with this result is that in practice
we have hash functions which are secure for apparently no obvious
reason. Some advancement in that problem can be made in redefining
what entitlements have the term \textquotedblleft{}fully described
functions\textquotedblright{}, because it appears that the term \textquotedblleft{}fully
described function\textquotedblright{} in ROM goes along lines from
figure 6. Consequently that assumption is false because we have the
case from figure 4. 

On the other hand ROM results show that if complexities of a function\textquoteright{}s
description and complexities of inputs are the same then the function
is indistinguishable from random oracle,, although that scenario is
dismissed on practicality grounds. Again, the algorithm from figure
4 clearly satisfies requirement of description and input complexity
being the same. Also calculating an instance of the algorithm from
figure 4 is more than practical. Following ROM argumentation we have
a deterministic process with output which can not be distinguished
from true randomness.

There may be some pointers for a design of hash functions and ciphers.
The complexity in cryptography is mainly acquired by focusing on complexity
of individual functions and functions predefined compositions (in
principle same concept as figure 5). The alternative may be to make
functions (f\textquoteright{}s and g\textquoteright{}s) simple but
to compose them in dynamical and unpredictable way (applying high
conditional complexity figure 4) therefore achieving requirement for
the same complexity of the function inputs and the full function description

\section{Conclusion}

The following is concluded:
\begin{enumerate}
\item An algorithm can definitely be considered as a form of a composite
function.
\item Giving the description of an algorithm does not mean giving a full
description of the composite function, but does mean that the algorithm
is deterministic.
\item There are some cases where the full function\textquoteright{}s description
complexity and the input complexity coincide while the underlying
algorithm is fairly simple and easy to execute.
\end{enumerate}
From an algorithm perspective, there is a formulation from Bohm and
Jacopini\textquoteright{}s work \citep{key-9}which demonstrates that
all programs could be written in terms of only three control structures:
(a) The sequence structure, (b) The selection structure and (c) The
repetition structure. It can be argued that formalism may not be relevant
when \textquoteleft{}(b) the selection structure\textquoteright{}
is used in an algorithm causing conditional complexity. In other words
we can have deterministic processes which may not be possible to formally
distinguish from the true randomness.


\begin{thebibliography}{9}
\bibitem[1]{key-1}http://www.ecrypt.eu.org/stream/mag.html

\bibitem[2]{key-3}http://www.literateprogramming.com/mccabe.pdf

\bibitem[3]{key-2}http://en.wikipedia.org/wiki/Collatz\_conjecture

\bibitem[4]{key-4}http://en.wikipedia.org/wiki/Kolmogorov\_complexity

\bibitem[5]{key-5}Craig Alan Feinstein: arXiv:math/0312309v16 {[}math.GM{]}

\bibitem[6]{key-6}http://www.wolframscience.com/thebook.html

\bibitem[7]{key-7}John von Neumann (quoted in Knuth, Vol.2)

\bibitem[8]{key-8}http://eprint.iacr.org/1998/011.pdf

\bibitem[9]{key-9}http://en.wikipedia.org/wiki/Structured\_program\_theorem
\end{thebibliography}
\end{document}